\documentclass[superscriptaddress,pra,10pt,aps,noeprint,twocolumn]{revtex4-1}
\usepackage[pdftex]{graphicx}
\usepackage{graphicx}

\usepackage[latin1]{inputenc}
\usepackage{amsmath}
\usepackage{ulem}
\usepackage{bbold}
\usepackage[T1]{fontenc}
\usepackage{color,bm, braket}
\usepackage[emulate = units]{siunitx}

\usepackage{xcolor}
\usepackage{upgreek}

\newcommand {\mo}{MoS$_2$}

\begin{document}
	\DeclareGraphicsExtensions{.pdf}
	
	\title{Comprehensive tunneling spectroscopy of \\quasi-freestanding MoS$_2$ on graphene on Ir(111)}
	
	\author{Clifford Murray}
	\email{murray@ph2.uni-koeln.de}
	\affiliation{II. Physikalisches Institut, Universit\"{a}t zu K\"{o}ln, Z\"{u}lpicher Stra\ss e 77, 50937 K\"{o}ln, Germany}
	\author{Wouter Jolie} 
	\affiliation{II. Physikalisches Institut, Universit\"{a}t zu K\"{o}ln, Z\"{u}lpicher Stra\ss e 77, 50937 K\"{o}ln, Germany}
	\affiliation{Institut f{\"ur} Materialphysik, Westf\"{a}lische Wilhelms-Universit\"{a}t M\"{u}nster, Wilhelm-Klemm-Stra{\ss}e 10, 48149 M\"{u}nster, Germany}
	\author{Jeison A. Fischer} 
	\affiliation{II. Physikalisches Institut, Universit\"{a}t zu K\"{o}ln, Z\"{u}lpicher Stra\ss e 77, 50937 K\"{o}ln, Germany}
	\author{Joshua Hall}
	\affiliation{II. Physikalisches Institut, Universit\"{a}t zu K\"{o}ln, Z\"{u}lpicher Stra\ss e 77, 50937 K\"{o}ln, Germany}
	\author{Camiel van Efferen}
	\affiliation{II. Physikalisches Institut, Universit\"{a}t zu K\"{o}ln, Z\"{u}lpicher Stra\ss e 77, 50937 K\"{o}ln, Germany}
	\author{Niels Ehlen}
	\affiliation{II. Physikalisches Institut, Universit\"{a}t zu K\"{o}ln, Z\"{u}lpicher Stra\ss e 77, 50937 K\"{o}ln, Germany}
	\author{Alexander Gr\"{u}neis}
	\affiliation{II. Physikalisches Institut, Universit\"{a}t zu K\"{o}ln, Z\"{u}lpicher Stra\ss e 77, 50937 K\"{o}ln, Germany}
	\author{Carsten Busse}
	\affiliation{II. Physikalisches Institut, Universit\"{a}t zu K\"{o}ln, Z\"{u}lpicher Stra\ss e 77, 50937 K\"{o}ln, Germany}
	\affiliation{Institut f{\"ur} Materialphysik, Westf\"{a}lische Wilhelms-Universit\"{a}t M\"{u}nster, Wilhelm-Klemm-Stra{\ss}e 10, 48149 M\"{u}nster, Germany}
	\affiliation{Department Physik, Universit\"{a}t Siegen, 57068 Siegen, Germany}
	\author{Thomas Michely}
	\affiliation{II. Physikalisches Institut, Universit\"{a}t zu K\"{o}ln, Z\"{u}lpicher Stra\ss e 77, 50937 K\"{o}ln, Germany}

	\date{\today}

	\begin{abstract}
		We apply scanning tunneling spectroscopy to determine the bandgaps of mono-, bi- and trilayer \mo~grown on a graphene single crystal on Ir(111). Besides the typical scanning tunneling spectroscopy at constant height, we employ two additional spectroscopic methods giving extra sensitivity and qualitative insight into the $k$-vector of the tunneling electrons. Employing this comprehensive set of spectroscopic methods in tandem, we deduce a bandgap of $2.53\pm0.08$\,eV for the monolayer. This is close to the predicted values for freestanding \mo~and larger than is measured for \mo~on other substrates. Through precise analysis of the `comprehensive' tunneling spectroscopy we also identify critical point energies in the mono- and bilayer \mo~band structures. These compare well with their calculated freestanding equivalents, evidencing the graphene/Ir(111) substrate as an excellent environment upon which to study the many feted electronic phenomena of monolayer \mo~and similar materials. Additionally, this investigation serves to expand the fledgling field of the comprehensive tunneling spectroscopy technique itself.
	\end{abstract}
	
	
	\maketitle
	
	\section{Introduction}	
		
	The various exciting properties of monolayer molybdenum disulfide (ML-\mo), the paradigmatic semiconducting transition metal dichalcogenide (TMDC), are well-documented \cite{Ganatra2014,Manzeli2017}. Amongst these its large, direct bandgap is promising for the electronics communities, and is a basic quality to be characterized. Large-scale flakes can be grown epitaxially \cite{Hall2017,Najmaei2013,Li2017} or exfoliated \cite{Novoselov2005,Radisavljevic2011}, but reliable characterization of the pristine electronic bandgap remains problematic.
	
	Optical measurements are influenced by the large exciton binding energy of ML-\mo. Standard angle-resolved photoemission spectroscopy (ARPES) has no access to the conduction band unless it is shifted below the Fermi energy $E_\text{F}$ through heavy doping. This, however, also leads to band distortion and bandgap renormalization due to the change in dielectric environment \cite{Miwa2015,Ehlen2018,Liang2015,Erben2018}. Pump-probe ARPES can measure the electronic bandgap \cite{Cabo2015}, but suffers from poor energy resolution.
	
	Scanning tunneling spectroscopy (STS) can directly access the electronic density of states above and below $E_\text{F}$, and it has indeed been performed on ML-\mo~on a variety of substrates. However, the substrates \textemdash~metallic by necessity \textemdash~tend to screen, gate, and/or mechanically strain the \mo. This leads to the predicted freestanding bandgap of $E_\text{g}\approx2.8$\,eV \cite{Cheiwchanchamnangij2012,Qiu2013,Ramasubramaniam2012,Shi2013} being considerably reduced. For example the bandgap measured by constant height STS is $E_\text{g}=1.74$\,eV on an Au substrate \cite{Bruix2016}, $2.01$\,eV on graphene/SiC \cite{Liu2016a}, $2.17$\,eV on quartz \cite{Rigosi2016}, $2.20$\,eV on graphene/Au \cite{Shi2016}, and variously $1.9$\,eV \cite{Lu2015}, $2.15$\,eV \cite{Zhang2014} or $2.40$\,eV \cite{Huang2015} on graphite. In addition to simply reducing the bandgap size, substrate coupling will affect each band differently \textemdash~due to the differing planar nature of the Mo and S orbitals, the band structure is distorted inhomogeneously across the \mo~Brillouin zone (BZ) \cite{Bruix2016}. Large bandgaps of $E_\text{g}\approx2.65$\,eV \cite{Hong2018} and $\approx2.7$\,eV \cite{Krane2016} have been reported, but only in locations where the ML-\mo~is locally decoupled from an inhomogeneous substrate. On top of all this, practical difficulties due to sulfur's relatively high vapour pressure had, until recently \cite{Hall2017}, hindered molecular beam epitaxy (MBE) synthesis of \mo. Thus close-to-freestanding \mo~flakes of sufficient size, quality, and cleanness on STS-permitting substrates have remained elusive. 
	
	Additional to the complications caused by the metallic substrates on which it is performed, there are shortcomings in the typical practice of STS. It has recently been shown by Zhang \textit{et al.} \cite{ZhangProbing2015} that constant height STS alone is insufficient for accurate bandgap determination, as states from the edge of the BZ can go undetected due to their reduced decay length. Therefore it remains an open question, how accurately the constant height STS measured bandgaps represent the magnitude of the ML-\mo~direct gap. In contrast, constant current STS allows the tip to move closer to the sample to give access to these weaker signals, while $\kappa$ mode STS (explained below) allows identification of the states' location within the BZ. 
	
	In this work we present high-quality ML-, bilayer \linebreak(BL-), and trilayer (TL-)\mo~which is well decoupled from its graphene/Ir(111) substrate. Following the approach of Zhang \textit{et al.} \cite{ZhangProbing2015}, we use `comprehensive STS' (constant height, constant current and $\kappa$ modes together) to identify not only the bandgaps but also various critical point energies (CPEs), i.e. local extrema in the band structure. These measured energies compare favourably with those of theoretical calculations for the freestanding materials, evidencing this system as an opportunity to study the inherent characteristics of mono- or few-layer \mo~without obtrusive substrate effects.
	
	Moreover, our analysis makes plain that standard constant height STS fails to detect both the valence band maximum and conduction band minimum, and thus does not measure the bandgap of ML-\mo. This has implications for the interpretation of STS data of ML-\mo, and indeed other materials with extremal points forming the bandgap at large parallel momenta. Comprehensive STS is not only more sensitive, but enables also the determination of the CPEs making up the tunneling spectrum. As shall be demonstrated here, this can prevent the false assignment of a band edge. It is thus a vital tool in the determination of the electronic structure of the semiconducting TMDCs. The technique and its associated analysis have only seen a few instances of usage \cite{ZhangProbing2015,Zhang2017a,Krane2018}, and so a broader implementation could be wished.
	
	\section{Methods}
	
	The sample is prepared \textit{in situ} at pressures $p<5\times10^{-10}$\,mbar. The Ir(111) single crystal is cleaned by Ar$^+$ ion sputtering and annealing at temperatures $T\approx1500$\,K. As described in Ref.\,\cite{Coraux2009}, a closed monolayer of graphene (Gr) is grown on Ir(111) via temperature programmed growth and chemical vapour deposition (CVD) at $T\approx1370$\,K. ML- to few-layer \mo~is subsequently grown on the Gr/Ir(111) substrate by van der Waals MBE, according to the methods developed in Ref.\,\cite{Hall2017}. Mo is evaporated from an e-beam evaporator and S from FeS$_2$ granules in a Knudsen cell. Specifically, we evaporate Mo in a S background pressure of $5\times10^{-9}$\,mbar onto the room temperature substrate, and then anneal the system to $1050$\,K in the same S background pressure. The process of co-evaporation then annealing can be repeated in cycles, in order to promote well-oriented, multiple-layer growth.
	
	Scanning tunneling microscopy (STM) and STS are performed at $T=5$\,K and $p<10^{-11}$\,mbar with a tungsten tip. For STS we use a lock-in amplifier with modulation frequency $777$\,Hz and modulation amplitudes $V_{\text{mod}}=4-8$\,mV$_{\text{rms}}$ \textemdash~together with thermal broadening this yields experimental resolution of ${\Delta}E\approx\sqrt{(3.3k_{\text{B}}T)^2+(2.5eV_{\text{mod}})^2}\approx20$\,meV or better \cite{Morgenstern2003}. We perform comprehensive STS comprised of three different modes: constant height (recording $(\text{d}I/\text{d}V)_Z$), constant current ($(\text{d}I/\text{d}V)_I$) and $\kappa$ ($(\text{d}I/\text{d}Z)_I$), where $I$ is the tunneling current, $V$ the bias voltage and $Z$ the tip-to-sample distance or `height'. The principles of these three modes shall be discussed. 
	
	For both constant height and constant current STS we measure the $\text{d}I/\text{d}V$ signal while $V$ is ramped, giving information on the local density of states of the sample \cite{Stroscio1986}. Though constant height STS allows both valence and conduction bands to be measured in a single spectrum, certain states may go undetected if $Z$ is too large. Constant current STS does not permit ramping across $E_\text{F}$ but offers greater dynamic range; the tip can move towards the sample and thereby detect some suppressed signals missed in constant height mode. This suppression can be due to the fact that a state with finite parallel momentum $k_\parallel$ will decay into the vacuum with an inverse decay length
	\begin{equation}\label{eq1}
	\kappa=[(2m\bar{\phi}/\hbar^2)+k_\parallel^2]^{1/2}\text{,}
	\end{equation} where $m$ is the free electron mass and $\bar{\phi}=(\phi_{t}+\phi_{s}-e|V|)/2$ is the bias-dependent tunneling barrier between tip and sample with work functions $\phi_{t}$ and $\phi_{s}$ respectively \cite{Feenstra1987,Tersoff1983}. Thus, states at the edge of the BZ decay more quickly into the vacuum than those at the center. This necessitates the tip moving closer to detect them, especially if stabilization was performed at a voltage (energy) where $\Gamma$-point states dominate.
	
	We indirectly measure $\kappa$ and thus $k_\parallel$ through $(\text{d}I/\text{d}Z)_I$ mode STS. Here the lock-in modulates the height ($Z_{\text{mod}}=4-8$\,pm) while $V$ is ramped at constant $I=I_0$ as before. Considering a tunneling current $I\propto{e^{-2{\kappa}Z}}$ \cite{Tersoff1985} one finds
	\begin{equation}\label{eq2}
	 \frac{\text{d}I}{\text{d}Z}\Big\rvert_{I_0}\propto{-2{\kappa}e^{-2{\kappa}Z}}=-2{\kappa}{I_0}\text{;}
	\end{equation}		
	we measure this and thereby extract an effective tunneling decay constant. Through comparison with the spectra obtained via the two other modes, one can assign features of the STS spectra to particular critical points in the BZ. Thus, a degree of $k$-space resolution has been added to the traditional STS. We note that inside the \mo~bandgap, when the tip moves very close to the sample, the `thick barrier' limit implicitly assumed in Eq.~(\ref{eq2}) does not necessarily hold and Gr states may contribute to the tunneling current. Therefore we do not draw inferences from $\kappa$ values within the bandgap.
	
	\section{Experimental Results}
	
	\begin{figure}[t]
		\centering
		\includegraphics[width=\columnwidth]{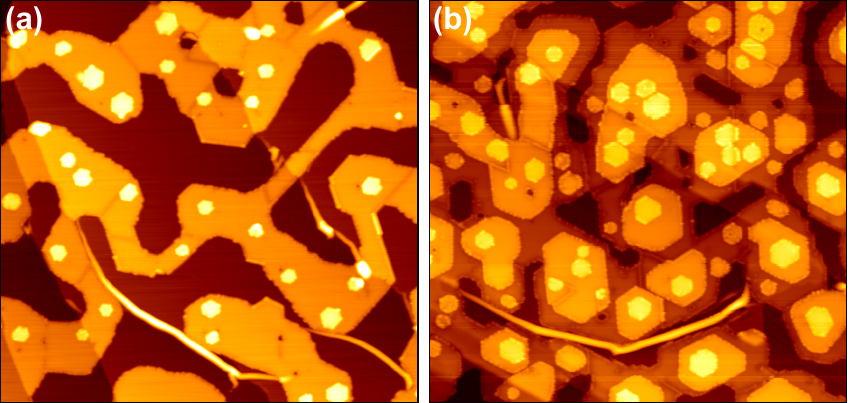}
		\caption{Constant current STM topographs of \mo~on Gr/Ir(111). (a) \mo~coverage of $0.6$ layers. (b) \mo~coverage of $1.4$ layers. Small areas of exposed Gr are visible. The TL forms islands of $\approx20$\,nm diameter. Gr wrinkles are visible in the lower section in both topographs. STM parameters: (a) $V=1.5$\,V, $I=0.01$\,nA; (b) $V=1.0$\,V, $I=0.08$\,nA; each image size $200\times190$\,nm$^\text{2}$.
			\label{fig1}}
	\end{figure}
	
	STM topographs of two typical \mo~samples are shown in Fig.~\ref{fig1}. In (a) an \mo~coverage of around $0.6$ layers yields a network of ML-\mo~extending over the Gr/Ir(111) substrate, crossing several Ir step edges. It is decorated by small BL islands of $\approx10$\,nm diameter. The cleanness and low defect density of the \mo, reported previously \cite{Hall2017}, were verified with STM here.  Grain boundaries are visible between ML flakes of different orientation. The majority of these are mirror twin boundaries (MTBs), the properties of which are discussed in Ref.\,\cite{Jolie}. In the lower section of the topograph a Gr wrinkle can be seen, resultant from the CVD growth.
	
	With a higher coverage of approximately $1.4$ layers, shown in Fig.~\ref{fig1}(b), the sample exhibits ML-, BL- and TL-\mo~islands in coexistence. Small areas of exposed Gr are visible below the nearly-closed ML. Large, well-oriented BL and $\approx20$\,nm diameter TL islands form on top. MTBs are seen to also occur in the BL.
	
	\subsection{Constant height STS of mono- and bilayer \mo}
	
	\begin{figure}[b]
		\centering
		\includegraphics[width=\columnwidth]{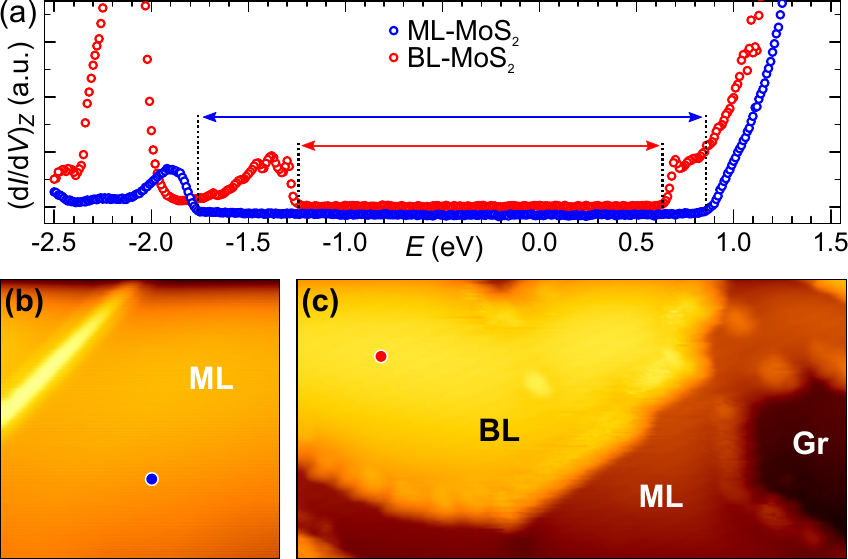}
		\caption{(a) Constant height spectra of ML- and BL-\mo~(in blue and red respectively). Assignment of the bandgaps based only on this STS method is shown. The spectra were taken at the points marked in topographs (b) and (c). The different areas of each sample are indicated for clarity. In (b) a MTB is seen in the top-left corner. STS/M parameters (with stabilization voltage $V_\text{st}$, stabilization current $I_\text{st}$): (a) $V_\text{st}=1.5$\,V; ML $I_\text{st}=0.1$\,nA, BL $I_\text{st}=1.0$\,nA; (b) $V=0.9$\,V, $I=0.10$\,nA, image size $10\times10$\,nm$^2$; (c) $V=1.0$\,V, $I=0.08$\,nA, image size $28\times14$\,nm$^2$.
			\label{fig2}}
	\end{figure}
	
	For illustrative purposes we first determine the bandgaps of ML- and BL-\mo~using constant height STS only, as is typically done in the literature for this and other TMDCs. Fig.~\ref{fig2}(a) shows two exemplary constant height spectra of ML- and BL-\mo. Topographs in Fig.~\ref{fig2}(b,c) show where the respective spectra were obtained. Note that all spectra in this work were recorded at locations at least $5$\,nm from any defects \textemdash~e.g. edges, MTBs, or point-defects \textemdash~to avoid any perturbation or confinement effects which these may cause.
	As is common in the literature, we here define the band edges to be where the d$I$/d$V$ signal becomes clearly discernible from background noise levels. Through this approach, we find the valence band maximum (VBM) to be located at $-1.77$\,eV and the conduction band minimum (CBM) to be at $0.86$\,eV for ML-\mo. Similarly for the BL, the corresponding band edges are found to be at $-1.24$\,eV and $0.63$\,eV. This would yield bandgap estimates of $2.63$\,eV and $1.87$\,eV for ML- and BL-\mo~respectively. However, it shall be demonstrated that these bandgap determinations for \mo~are unreliable.
	
\begin{figure}[h]
		\centering
		\includegraphics[width=\columnwidth]{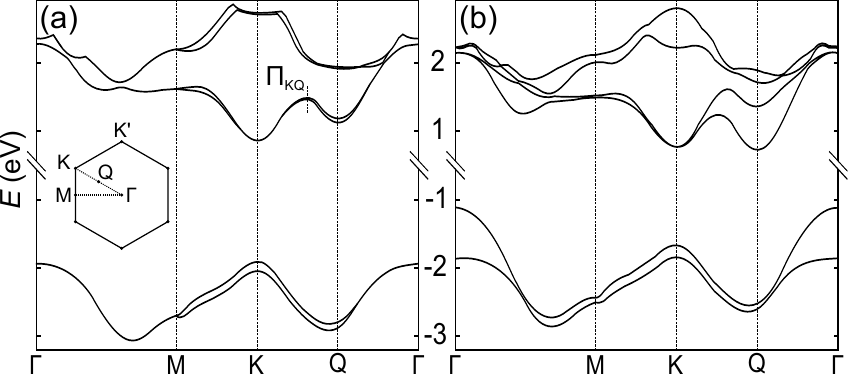}
		\caption{Sketched band structures of freestanding (a) ML- and (b) BL-\mo. Redrawn after Ref.\,\cite{Cheiwchanchamnangij2012} and adapted to reflect comparison with other DFT calculations \cite{Ramasubramaniam2012,Qiu2013,Ramasubramaniam2011}; the figure should serve only as a generic outline. The first BZ is shown as an inset, with the high-symmetry points and Q-point marked. Also indicated is a local maximum between the K and Q-points in the ML CB, labeled here $\Pi_{\text{KQ}}$. For ease of reference the entire band structures have been rigidly shifted to approximately match our energies, rather than fixing $0$\,eV at the VBM as is typical in DFT.
			\label{fig3}}
	\end{figure}

	We briefly consider the band structures of ML- and BL-\mo~close to $E_\text{F}$, to guide proceeding STS analysis. The band structures sketched in Fig.~\ref{fig3} are based on previous density functional theory (DFT) calculations \cite{Cheiwchanchamnangij2012,Ramasubramaniam2012,Qiu2013,Ramasubramaniam2011}. As seen in Fig.~\ref{fig3}(a), the ML has a direct bandgap located at the K-point. The VB is split by $\approx145$\,meV at K due to spin-orbit coupling, and a maximum at $\Gamma$ lies close in energy \cite{Cheiwchanchamnangij2012,Qiu2013,Ramasubramaniam2012,Zhu2011,Ehlen2018}. In contrast, the BL (b) has a smaller and indirect bandgap, with the VBM located at the $\Gamma$-point and the CBM at the Q-point. The critical points at K and Q in the CB lie close in energy however, and so the true location of the CBM is debated in the literature \cite{Du2018,LiuGB2015}. The VB is split at the $\Gamma$-point due to interlayer hopping \cite{Debbichi2014}. 
	
	\subsection{Comprehensive STS of monolayer \mo}
	
	\begin{figure*}[]
		\centering
		\includegraphics[width=0.7\textwidth]{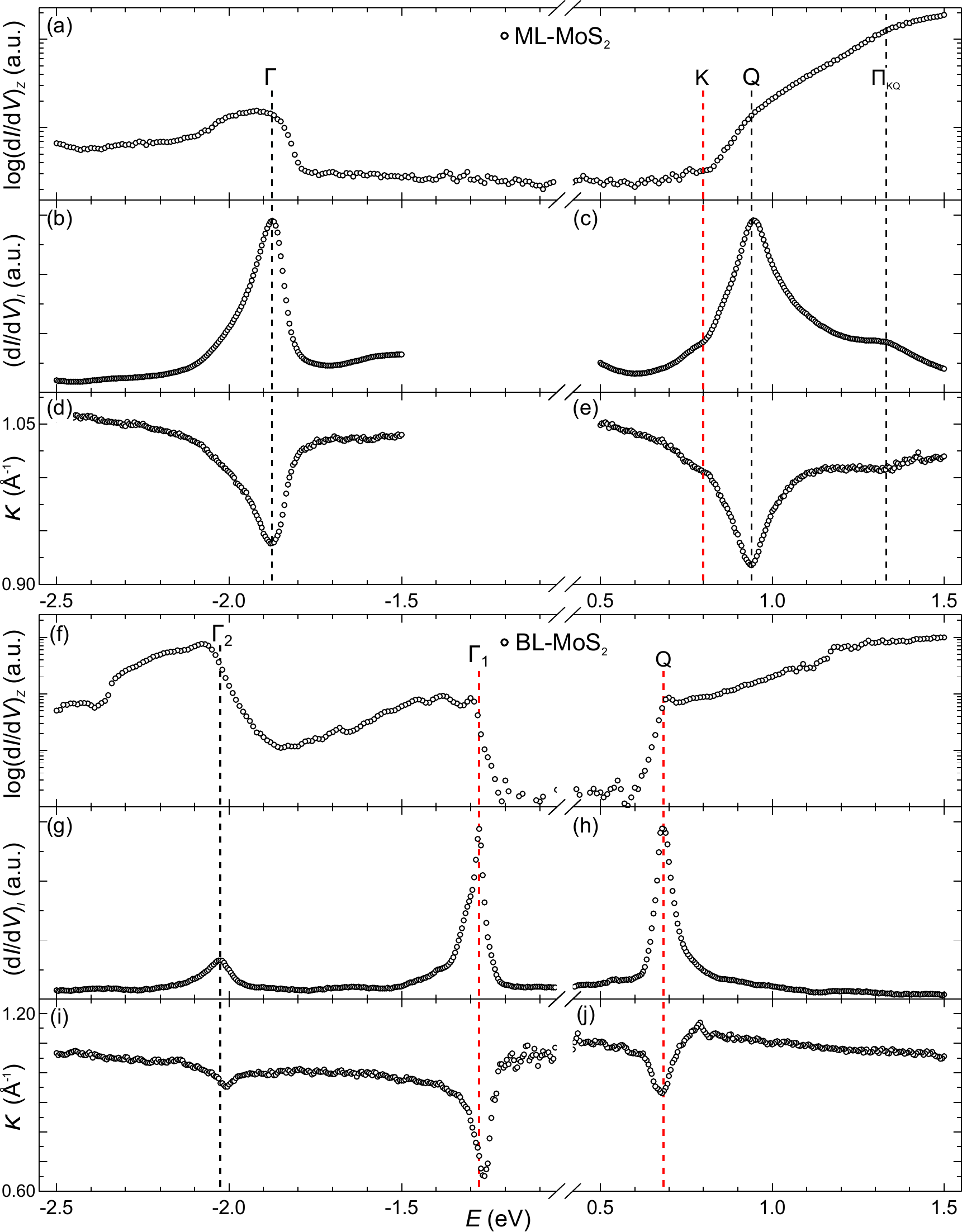}
		\parbox{\textwidth}{\caption{Comprehensive STS of (a-e) ML- and (f-j) BL-\mo. The spectra were obtained at the locations shown in Fig.~\ref{fig2}(b,c); note that Fig.~\ref{fig4}(a,f) show the same spectra as Fig.~\ref{fig2}(a), here plotted on a logarithmic intensity scale. (a,f) Constant height $(\text{d}I/\text{d}V)_Z$ STS spectra. (b,c,g,h) Constant current $(\text{d}I/\text{d}V)_I$ STS spectra performed over the VB and CB edges of the respective systems. (d,e,i,j) $\kappa$ (recording $(\text{d}I/\text{d}Z)_I$) STS spectra performed over the VB and CB edges. (a-j) Assigned critical point energies are marked by dashed black lines; those critical points which constitute a VBM or CBM are dashed red. STS parameters: (a,c,e,f,h,j) $V_\text{st}=1.5$\,V, (b,d,g,i) $V_\text{st}=-2.5$\,V; (a-e) $I_\text{st}=0.10$\,nA, (f) $I_\text{st}=1.00$\,nA, (g-j) $I_\text{st}=0.05$\,nA.}
			\label{fig4}}
	\end{figure*}

	In Fig.~\ref{fig4} exemplary sets of comprehensive STS on ML- and BL-\mo~ are shown. The three different STS modes are considered together and for both the ML and BL are compared across at least five sets of spectra, taken on various islands and with different STM tips. Through this, some critical points in the respective band structures can be assigned. 
	
	Beginning with the VB of the ML, Fig.~\ref{fig4}(a) shows the same constant height spectrum as in Fig.~\ref{fig2}(a) now plotted logarithmically. In (b) constant current STS yields a main peak at $-1.87$\,eV with a slight shoulder towards larger binding energies. In (d) the corresponding $\kappa$ measurement shows a dip, also at $-1.87$\,eV. This dip to $0.93\text{\,\AA}^{-1}$ indicates a sudden drop in the effective tunneling decay constant of the states there, i.e. states with less $k_\parallel$. (A discussion of the actual $k_\parallel$ values extracted from $\kappa$ follows in Sec.~\ref{Discussion}.) Considering the drop in $\kappa$ and with reference to the band structure of ML-\mo~[Fig.~\ref{fig3}~(a)], we must assume this feature to be due to the $\Gamma$-point. Though the VBM is expected to be the upper of the spin-split bands at the K-point, we can expect the spectrum to be totally dominated by states from $\Gamma$. Firstly, the states at K decay faster into the vacuum due to their high $k_\parallel$. Furthermore, calculation has shown that the orbital character at the $\Gamma$-point is predominantly Mo-d$_{z^2}$, while at K it is predominantly Mo-d$_{xy}$,d$_{{x^2}-{y^2}}$, i.e. mostly out-of-plane and mostly in-plane respectively \cite{Komsa2013a,Cappelluti2013,Padilha2014}. Thus, if the bands at K and $\Gamma$ lie sufficiently close in energy we would expect the $\Gamma$-band to mask the K-band in our STS signal. Indeed in \textit{ab initio} calculations the separation between the upper K-band (K$_\uparrow$) and the band at $\Gamma$ is found variously to be around ${\Delta}{\Gamma}\text{K}_{\uparrow}=0.04$ \cite{Cheiwchanchamnangij2012}, $0.05$ \cite{Zhu2011}, $0.12$ \cite{Ramasubramaniam2012} or $0.19$\,eV \cite{Qiu2013}. We conclude that both branches of the spin-split band at K are masked by $\Gamma$ states. An estimate for the position of K$_\uparrow$ (i.e. the VBM) can nonetheless be made. In ARPES experiments on ML-\mo~grown on Gr/Ir(111) by the same method as in this work, a separation between $\Gamma$ and K$_\uparrow$ of ${\Delta}{\Gamma}\text{K}_{\uparrow}=0.11$\,eV was found \cite{Ehlen2018,Ehlen}. This energy separation would locate K$_\uparrow$ at $-1.76$\,eV in our case. We consider lower and upper bounds based on the aforementioned DFT calculations to be ${\Delta}{\Gamma}\text{K}_{\uparrow}=0.04$\,eV \cite{Cheiwchanchamnangij2012} and ${\Delta}{\Gamma}\text{K}_{\uparrow}=0.19$\,eV \cite{Qiu2013} respectively, i.e. for K$_\uparrow$ to lie between $-1.83$\,eV and $-1.68$\,eV. Taking these bounds as a conservative uncertainty, we estimate the VBM of our ML-\mo~system to be located at $-1.76\pm0.08$\,eV.
	
	The CB of ML-\mo~also shows various features in constant current STS, Fig.~\ref{fig4}(c). A main peak at $0.94$\,eV is flanked by a small shoulder at $0.80$\,eV and, towards higher energies, a broad hump at $1.32$\,eV. In (e), $\kappa$ shows a clear dip to $0.91\text{\,\AA}^{-1}$ at $0.94$\,eV. Consulting the theoretical band structure [Fig.~\ref{fig3}(a)], a local minimum close to the CB edge is expected at the Q-point, and any $\Gamma$ states are much further from the Fermi level \textemdash~thus this feature must be assigned to the Q-point. Across our sets this peak tended to take one of two values \textemdash~either $0.86$\,eV or $0.95$\,eV approximately \textemdash~and typically has a broad shape suggestive of more than one contributing state. We find no correlation of the Q-point peak value to the lateral position in the \mo~layer. The properties could be due to the spin-splitting of the band at the Q-point, predicted to be of magnitude $0.03 - 0.08$\,eV  \cite{Qiu2013,Ramasubramaniam2012,Cheiwchanchamnangij2012,Zhu2011}. The faint shoulder at $0.80$\,eV has no obvious corresponding feature in (e) here, though a small peak was occasionally seen at this energy in $\kappa$ spectra. The feature was practically undetectable in constant height STS, suggesting that it originates from states of large $k_\parallel$ and/or of mostly in-plane orbital nature. This fact, combined with a small peak sometimes seen in $\kappa$ and with consultation of the ML-\mo~band structure, compels assigning this feature to states at the K-point. This represents the CBM of ML-\mo, found at $0.77\pm0.02$\,eV across the measured sets. This K-point extremum being detectable, in contrast to the K-point of the VB, can be explained by its orbital character. The K-point at the CBM is dominated by out-of-plane Mo-d$_{z^2}$ orbitals; at the VBM it is dominated by in-plane Mo-d$_{xy}$,d$_{{x^2}-{y^2}}$ orbitals \cite{Komsa2013a,Cappelluti2013,Padilha2014}. Finally, we assign the broad hump at $1.32$\,eV to the local maximum lying roughly halfway between the K- and Q-points, which we term $\Pi_{\text{KQ}}$. The (average) assigned CPEs for the ML are summarized in Table \ref{tbl:CPEsML}.
	
	\begingroup
	\squeezetable
	\begin{table}[h]
		\begin{ruledtabular}
			\caption{CPEs (eV) identified in ML-\mo~using comprehensive STS averaged over multiple sets, and the estimated CPE at K$_\uparrow$.}
			\label{tbl:CPEsML}
			\begin{tabular}{ccccc}
				$\Gamma$ & (K$_\uparrow$) & K & Q & $\Pi_{\text{KQ}}$ \\ 
				\colrule
				$-1.87\pm0.02$ & ($-1.76\pm0.08$) & $0.77\pm0.02$ & $0.90\pm0.05$ & $1.30\pm0.02$ \\ 
			\end{tabular}
		\end{ruledtabular}	
	\end{table}
	\endgroup
	
	It should be noted that the CPEs constituting band edges have alternatively been defined by Zhang \textit{et al.}~\cite{ZhangProbing2015} to be at the midpoint of the transition from TMDC to substrate states in the STS signal. This is practically equivalent to us taking the energy at FWHM of the peaks closest to $E_\text{F}$ \textemdash~for example in Fig.\ref{fig4}(c), with Gaussians fitted to the various features including the K-point shoulder, this would yield a CBM at $0.75$\,eV rather than $0.80$\,eV. However, due to ambiguities of peak-fitting in our spectra and for simplicity, we chose instead to define the band edges at the peak centers in constant current STS.
			
	\subsection{Comprehensive STS of bilayer \mo}
	
	In constant height STS of the BL-\mo, Fig.~\ref{fig4}(f), two sharp rises in intensity are seen in the VB. These are accompanied by clear peaks in constant current and clear dips in $\kappa$ measurements, (g) and (i) respectively. Based on their nature and with consideration of the generic BL band structure [Fig.~\ref{fig3}(b)], we can confidently assign the features marked at $-1.28$\,eV and $-2.03$\,eV to the split bands at the $\Gamma$-point, labelled $\Gamma_1$ and $\Gamma_2$ respectively. The signal from $\Gamma_2$ is much weaker in constant current and $\kappa$ STS because in the midst of the VB states the feedback loop has taken the tip further away from the sample, making it less sensitive to the onset of the $\Gamma_2$-band. Also for this reason, and considering their faster decay into the vacuum, it is wholly unsurprising that the K-point states expected close to $-2$\,eV were not reliably detected.
	
	Similarly in the CB a sharp rise in constant height [Fig.~\ref{fig4}(f)] coincides with a peak in constant current (h) and dip in $\kappa$ (j). The last of these indicates states from near the centre of the BZ. Consultation of Fig.~\ref{fig3}(b) shows that $\Gamma$-states lie deep in the CB, and thus we assign this feature at $0.68$\,eV to the Q-point representing the CBM. The small peak in $\kappa$ at around $0.79$\,eV in (j) could possibly be due to the K-point minimum, but this feature was not observed consistently enough with different tips to allow an unambiguous deconvolution. Considering the Q-point states' energetic proximity, their location at the band edge, and their smaller $k_\parallel$, they could be expected to mask the K-point states in STS. Indeed this issue is non-trivial; there is debate in the literature as to whether the CBM of BL-\mo~lies at the K-point \cite{Chu2015,Cheiwchanchamnangij2012} or at the Q-point \cite{Du2018,Mak2010,Debbichi2014}, a matter of relevance due to the lack of symmetry at the latter. The (average) assigned CPEs for the BL are summarized in Table \ref{tbl:CPEsBL}.
	
	\begin{table}[h]
		\begin{ruledtabular}
			\caption{CPEs (eV) identified in BL-\mo~using comprehensive STS averaged over multiple sets.}
			\label{tbl:CPEsBL}
			\begin{tabular}{ccc}
				${\Gamma}_2$  &  ${\Gamma}_1$ & Q  \\ 
				\colrule
				$-2.05\pm0.04$ & $-1.27\pm0.04$ & $0.69\pm0.03$ \\ 
			\end{tabular}
		\end{ruledtabular}	
	\end{table}
	
	With the band extrema identified in Tables \ref{tbl:CPEsML} and \ref{tbl:CPEsBL} we determine bandgaps of $E_\text{g}^{\text{ML}}=2.53\pm0.08$\,eV and $E_\text{g}^{\text{BL}}=1.96\pm0.05$\,eV. For the specific sets shown in Fig.~\ref{fig4} the bandgaps are $2.56$\,eV and $1.96$\,eV respectively.
	
	\begin{figure*}[]
		\centering
		\includegraphics[width=0.7\textwidth]{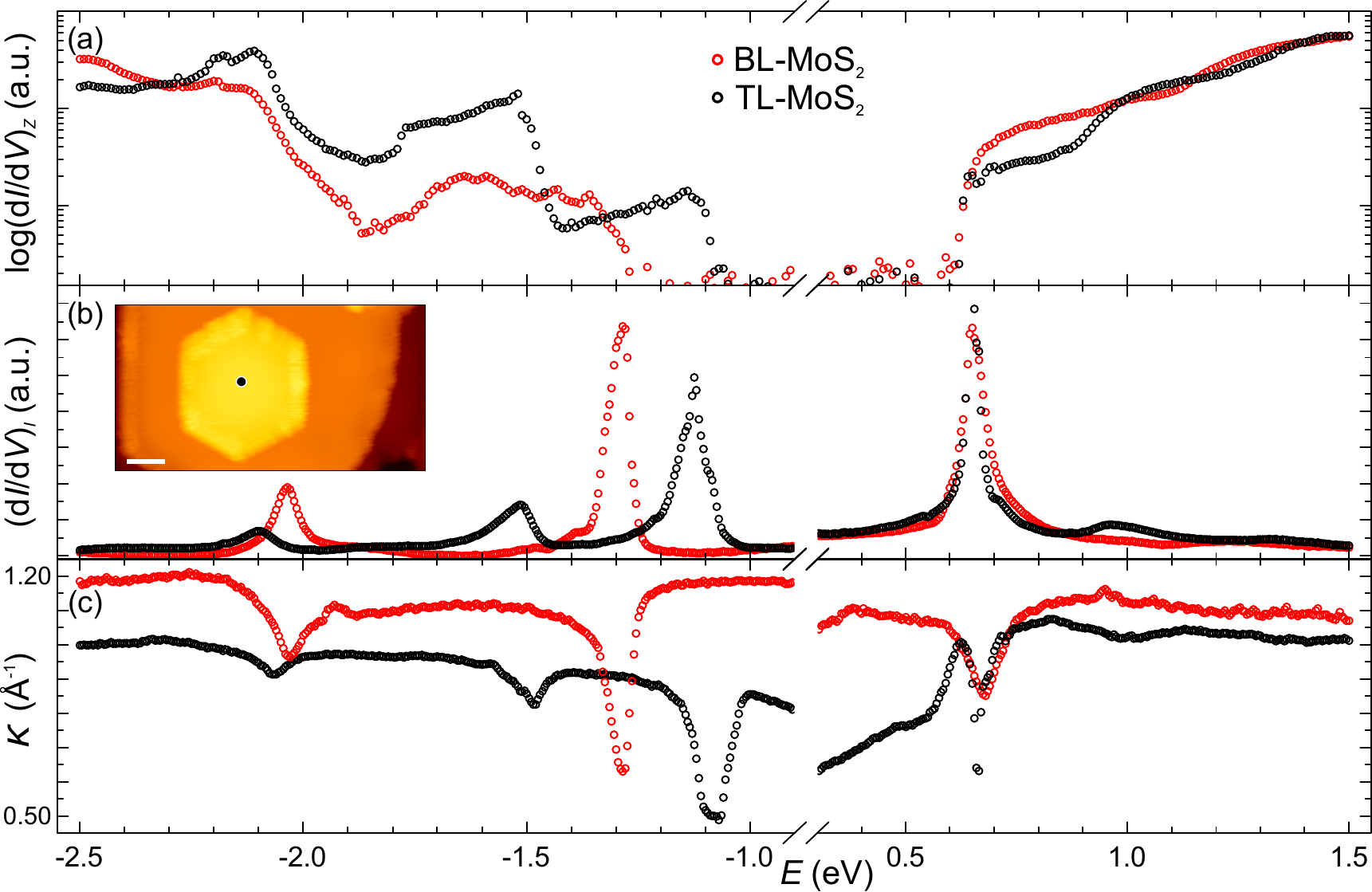}
		\parbox{\textwidth}{\caption{Comprehensive STS of TL-\mo~(black), with BL-\mo~(red) for comparison. The TL spectra were obtained at the location shown in the inset STM topograph; the BL spectra elsewhere on the same sample. Note that this is a different BL set from that shown in Fig.~\ref{fig2} and Fig.~\ref{fig4}, to give an idea of deviation within the spectroscopic data. STS parameters: (a) $V_\text{st}=1.5$\,V, $I_\text{st}=1.0$\,nA; (b,c) $V_\text{st}=-2.5$\,V and $1.5$\,V (for VB and CB respectively), $I_\text{st}=0.1$\,nA;	TL and BL spectra both obtained with the same parameters. STM (inset): $V=1.0$\,V, $I=0.07$\,nA, scale bar $5$\,nm.}
			\label{fig5}}
	\end{figure*}
	
	\subsection{Comprehensive STS of trilayer \mo}
	Fig.~\ref{fig5} shows comprehensive STS of TL-\mo, together with a BL set for comparison. The small size of the TL islands \textemdash~for example $\approx15$\,nm diameter in Fig.~\ref{fig5} \textemdash~means that interfering quantum confinement effects cannot be ruled out. Nonetheless, some qualitative features are obvious from the spectroscopic data. Namely, a third branch in the VB has appeared due to further splitting of the band at $\Gamma$, while the CB edge remains mostly unchanged, in line with theoretical calculations \cite{Padilha2014,Trainer2017}. This is discussed further in Sec.~\ref{Discussion}. A bandgap of $E_\text{g}^{\text{TL}}=1.77$\,eV is estimated based on Fig.~\ref{fig5}, though we provide this value tentatively due to limited statistics. Additionally, although investigations of such islands did not show lateral confinement, the aforementioned quantum-size issue should be noted.
			
	\section{Discussion}\label{Discussion}
	
	We find that using constant height STS alone would lead to a $0.10$\,eV overestimation of $E_\text{g}^{\text{ML}}$ (when compared with comprehensive STS analysis) because the measured states are not actually those at the respective band edges. In constant height STS both band extrema go undetected. In comprehensive STS the CBM at K is detected. The VBM is not detected but, importantly, a false assignment of the VBM is prevented through $\kappa$ measurements. One could wrongly assume that the peak in Fig.~\ref{fig4}(a,b) is due to the VBM (at the K-point), but the drop in $\kappa$ rules this out. Put simply: comprehensive STS sees more states, and when it is blind to certain states then it can tell us that this is the case.
	
	The benefits of the more thorough technique are further illustrated by accurate observation of layer-dependent phenomena in the \mo. It is known from the literature that the bandgap reduction with increasing thickness is due to the VBM \textemdash~specifically the $\Gamma$-point \textemdash~shifting to smaller binding energies, while the CBM does not change significantly in energy \cite{Padilha2014,Trainer2017,Bradley2015}. Using only constant height STS the CBM appears to shift by $0.23$\,eV towards $E_\text{F}$ upon addition of a second \mo~layer, whereas the shift is indeed much less drastic ($0.08$\,eV) in comprehensive STS. The continuation of these trends \textemdash~a static CBM and an up-shifting VBM \textemdash~is visible as the thickness is increased from BL to TL [Fig.~\ref{fig5}]. Additionally, the well-documented lifting of degeneracy in the $\Gamma$-band and its consequent splitting from one (ML) to two (BL) to three (TL) branches is clearly visible across the data sets. Thereby the coupling of each newly added \mo~layer to those underneath is seen through comprehensive STS. 
	
	The technique has its limitations, of course. As discussed, we could not unambiguously detect the K-point states which represent the VBM of ML-\mo, presumably due to their short decay length, in-plane orbital character, and proximity to the dominating $\Gamma$ point. States being hidden due to a combination of these factors is an issue; previous comprehensive STS investigations of ML-\mo~have also failed to identify the VBM \cite{Krane2018}. The K-point VBM was clearly detected in ML-MoSe$_2$ and ML-WSe$_2$, presumably because in these cases it is separated from the $\Gamma$-point by large energies of $0.39$\,eV and $0.64$\,eV respectively \cite{ZhangProbing2015}. However, the CBs of these materials and their sulfide analogues exhibit a trend \textemdash~the K-point STS signal becomes less and less prominent as it moves energetically closer to the Q-point \cite{ZhangProbing2015}.
	
	Measuring $\kappa$ helps reveal a state's location in the BZ, but extracting the corresponding values of $k_\parallel$ proves non-trivial. In Eq.~(\ref{eq1}) the only unknown variable is the energy barrier $(\phi_{t}+\phi_{s})/2$. We can set this (to $2.5$\,eV) to obtain reasonable $k_\parallel$ values for most CPEs. However, this is an \textit{ad hoc} adjustment and it fails for some CPEs regardless. Similar problems arise in $\kappa$ measurements in the literature \cite{ZhangProbing2015,Krane2018}. The values of $\kappa$ given here remain valid; we additionally take $I(Z)$ spectra at various bias voltages, to which $I(Z)\propto{e^{-2{\kappa}Z}}$ is then fitted, showing excellent agreement with $(\text{d}I/\text{d}Z)_I$ spectra. We suggest the difficulty in translating $\kappa$ into actual $k_\parallel$ values is due to an oversimplified picture of the tunneling that forms the basis of Eq.~(\ref{eq1}). Nonetheless, $\kappa$ serves as a useful qualitative measure of a state's position in the BZ relative to states energetically nearby.
	
	A more puzzling issue is an apparent mismatch between STS and ARPES studies. Specifically, the $\Gamma$-point in the VBM of ML-\mo~on Gr/Ir(111) is found to be $-1.87$\,eV in STS (this work) but $-1.61$\,eV in ARPES \cite{Ehlen2018,Ehlen}. The $\Gamma$-points in BL-\mo~on the same substrate coincide however; $-2.05$\,eV and $-1.27$\,eV in STS (this work) compared with $-1.98$\,eV and $-1.33$\,eV in ARPES \cite{Ehlen2018,Ehlen}. In collaborative STS \cite{Bruix2016} and ARPES \cite{Miwa2015} investigation of ML-\mo~on Au(111), discrepancies of $0.10$\,eV and $0.15$\,eV were found for the $\Gamma$- and K-point respectively. Comprehensive STS on the same system showed further disagreement \cite{Krane2018}. A comparative study of comprehensive STS and ARPES (performed on the same sample in the same UHV chamber) would present a considerable experimental challenge, but would be worthwhile if the community is to address these problems of inconsistency.
		
	Despite the discussed experimental uncertainties, it is clear that ML-\mo~on Gr/Ir(111) is a very well decoupled system. Our conservative estimate $E_\text{g}^{\text{ML}}=2.53\pm0.08$\,eV represents the largest STS-measured bandgap of ML-\mo~on a homogeneous substrate. This value comes much closer to the freestanding $E_\text{g}\approx2.8$\,eV predicted by DFT \cite{Cheiwchanchamnangij2012,Qiu2013,Ramasubramaniam2012,Shi2013} than those of ML-\mo~measured on other substrates, such as graphite ($E_\text{g}=2.40$\,eV \cite{Huang2015}). If we would instead take the bandgap size of $2.63$\,eV based on constant height STS alone \textemdash~as is done in the literature with which we compare this work \textemdash~our system appears even better decoupled. ML-\mo~nanopatches suspended over Au(111) vacancy islands of roughly $3$\,nm diameter have shown an apparent bandgap of $\approx2.7$\,eV indicating that they are quasi-freestanding \cite{Krane2016}, but their small size leaves them liable to lateral quantum confinement effects. An apparent bandgap of $\approx2.65$\,eV has been reported for water-intercalated areas of ML-\mo~on graphite \cite{Hong2018}. However, the interficial water layer and defects resultant from the wet transfer process have competing doping effects and leave the \mo~inhomogeneous.
	
	The freestanding nature of our system is further apparent upon closer examination of the measured CPE values. In Table~\ref{tbl:CPEvsTheory} the energy separations of ML- and BL-\mo~CPEs can be compared with those of various DFT calculations. Taking into account that there is considerable discrepancy within the DFT results themselves, the measured CPEs agree reasonably with calculation. The bandgap $E_\text{g}^{\text{BL}}=1.96\pm0.04$\,eV also compares well with values $1.89$\,eV \cite{Cheiwchanchamnangij2012} and $1.83$\,eV \cite{Debbichi2014} from the literature.	
	
	Previous experiments on ML-\mo~on Gr/Ir(111) suggest weak substrate interaction also. \mo~islands are mobile enough to be moved laterally on the surface using the STM tip \cite{Hall2017}. Additional evidence of weak interaction was seen in photoluminescence spectroscopy, x-ray photoemission spectroscopy, temperature dependent Raman spectroscopy, and ARPES \cite{Ehlen2018}. For example, comparing Raman measurements at room temperature and at $4$\,K showed that the ML-\mo~does not follow the thermal expansion of its substrate. Instead its expansion resembles that of a freestanding layer, meaning that it is not strained by the substrate. In ARPES, no hybridization of Gr and \mo~bands was seen \cite{Ehlen2018}.
		
	\begin{table}[]
		\begin{ruledtabular}
			\caption{Comparison of CPE separations measured in ML- and BL-\mo/Gr/Ir(111) here with those of freestanding \mo~as calculated by various DFT approaches. In the ${\Delta}\text{KQ}$ column the energy of both spin orientations in the split band at Q are given.}
			\label{tbl:CPEvsTheory}
			\begin{tabular}{c|ccc}
				& ML CB & ML CB & BL VB \\ 
				Ref. & ${\Delta}\text{KQ}$ (eV) & ${\Delta}\text{K}\Pi_{\text{KQ}}$ (eV) & ${\Delta}\Gamma_{1,2}$ (eV) \\ \hline
				this work & $0.13\pm0.05$ & $0.53\pm0.03$ & $0.78\pm0.06$ \\ 
				\cite{Cheiwchanchamnangij2012} & $_\uparrow0.44$, $_\downarrow0.51$ & $0.71$ & $0.75$ \\
				\cite{Qiu2013} & $_\uparrow0.19$, $_\downarrow0.25$ & $0.48$ & - \\
				\cite{Ramasubramaniam2012,Ramasubramaniam2011} & $_\uparrow0.08$, $_\downarrow0.12$ & $0.30$ & $0.69$ \\		
				\cite{Zhu2011} & $_\uparrow0.13$, $_\downarrow0.17$ & $0.27$ & - \\
				\cite{Debbichi2014} & - & - & $0.76$ \\
			\end{tabular}
		\end{ruledtabular}	
	\end{table}
		
	\section{Conclusion}
	
	We have characterized the electronic structure of quasi-freestanding ML-, BL- and TL-\mo~on Gr/Ir(111) with high-precision STS analysis, whereby the bandgaps have been determined, various CPEs close to $E_\text{F}$ identified, and layer-dependent phenomena observed. The measured bandgap sizes are close to those of the freestanding material, showing that \mo~is well decoupled from this substrate. The measured CPEs can be cross-referenced with those predicted by DFT calculations from the literature, further corroborating this. Thus Gr/Ir(111) represents a substrate for STS investigations of the inherent properties of 2D-TMDCs, with minimal interference from gating, band-rehybridization, or strain effects. 
	
	This work implores the use of comprehensive STS where possible. The technique gives access to states otherwise undetectable, for example the CBM of ML-\mo~here. Moreover, it adds a degree of $k$-space resolution, allowing identification of band structure features and preventing false assignments, for example of the VBM of ML-\mo~here. Thus the supplementary constant current and $\kappa$ STS modes are crucial for accurately determining the bandgap of ML-\mo, or of similar semiconductors with band edges located near the BZ boundary.
	
	\begin{acknowledgments}
		This work was funded by the Deutsche Forschungsgemeinschaft (DFG, German Research Foundation) - Project number 277146847 - CRC 1238 (subprojects A01 and B06). W.J. acknowledges financial support from the Bonn-Cologne Graduate School of Physics and Astronomy (BCGS).
	\end{acknowledgments}

	\bibliography{./library}
	
\end{document}